\documentclass[prr,twocolumn,superscriptaddress,preprintnumbers,a4paper,amsmath,amssymb,showpacs,floatfix]{revtex4}
\usepackage{graphics}
\usepackage{graphicx}
\usepackage{bm}

\newcommand{\UPI}{U$_2$Pd$_2$In}

\newcommand{\utx}{U$_2$T$_2$X}

\begin{document}


\title{Noncollinear magnetic structure in \UPI~at high magnetic fields}

\author{K.~Proke\v{s}}
\email{prokes@helmholtz-berlin.de}
\affiliation{Helmholtz-Zentrum
Berlin f\"{u}r Materialien und Energie, Hahn-Meitner Platz 1, 14109 Berlin, 
Germany}

\author{M. Bartkowiak}
\affiliation{Helmholtz-Zentrum
Berlin f\"{u}r Materialien und Energie, Hahn-Meitner Platz 1, 14109 Berlin, 
Germany}

\author{D. I. Gorbunov}
\affiliation{Hochfeld-Magnetlabor Dresden (HLD-EMFL), 
Helmholtz-Zentrum Dresden-Rossendorf, 01328 Dresden, Germany}

\author{O. Prokhnenko}
\affiliation{Helmholtz-Zentrum
Berlin f\"{u}r Materialien und Energie,  Hahn-Meitner Platz 1, 14109 Berlin, 
Germany}

\author{O. Rivin}
\affiliation{Helmholtz-Zentrum
Berlin f\"{u}r Materialien und Energie,  Hahn-Meitner Platz 1, 14109 Berlin, 
Germany}
\affiliation{Physics Department, Nuclear Research Centre − Negev, 84190 
Beer-Sheva, Israel}

\author{P. Smeibidl}
\affiliation{Helmholtz-Zentrum
Berlin f\"{u}r Materialien und Energie, Hahn-Meitner Platz 1, 14109
Berlin, Germany}

\date{\today}
\pacs{74.70.Xa 61.50.Ks 74.25.-q 75.30.-m}
\begin{abstract}

We report an unexpected magnetic-field-driven magnetic structure in the 5$f$-electron Shastry-Sutherland system \UPI. This phase develops at low temperatures from a noncollinear antiferromagnetic ground state above the critical field of 25.8 T applied along the $a$-axis. All U moments have a net magnetic moment in the direction of the applied field, described by a ferromagnetic propagation vector $q_{F}$ = (0 0 0) and an antiferromagnetic component described by a propagation vector $q_{AF}$ = (0 0.30 $\frac{1}{2}$) due to a modulation in the direction perpendicular to the applied field. We conclude that this surprising noncollinear magnetic structure is due to a competition between the single-ion anisotropy trying to keep moments, similar to the ground state, along the [110]-type directions, Dzyaloshinskii-Moryia interaction forcing them to be perpendicular to each other and application of the external magnetic field attempting to align them along the field direction.

\end{abstract}

\maketitle
 

Magnetic frustration, competition between two or more exchange interactions and Dzyaloshinskii-Moryia interaction (DMI) commonly leads either to formation of strongly reduced magnetic moments, canted magnetic structures or even more exotic magnetic states such as spin ice and spin liquids~\cite{UPI99,UPI55,UPI56,UPI71,UPI72,UPI73}. Shastry-Sutherland (SS) model (and related models), which is exactly solvable~\cite{UPI62}, is well known to be a playground for studies of frustrated magnets. Nevertheless, only a handful of systems are known to adopt this type of lattice. The SrCu$_{2}$(BO$_{3}$)$_{2}$~\cite{UPI57,UPI58,UPI60} that is nonmagnetic at low temperatures in zero field, is the most widely studied material. Application of magnetic field, pressure and low-level doping lead to novel magnetic phases studied frequently by means of high-field and/or neutron scattering techniques. It has been shown that under such perturbations SrCu$_{2}$(BO$_{3}$)$_{2}$ exhibits a magnetic order. TmB$_{4}$ and GdB$_{4}$~\cite{UPI20,UPI61} can be named as other two examples of experimental realization of the SS lattice.

\begin{figure}
\includegraphics*[scale=0.2]{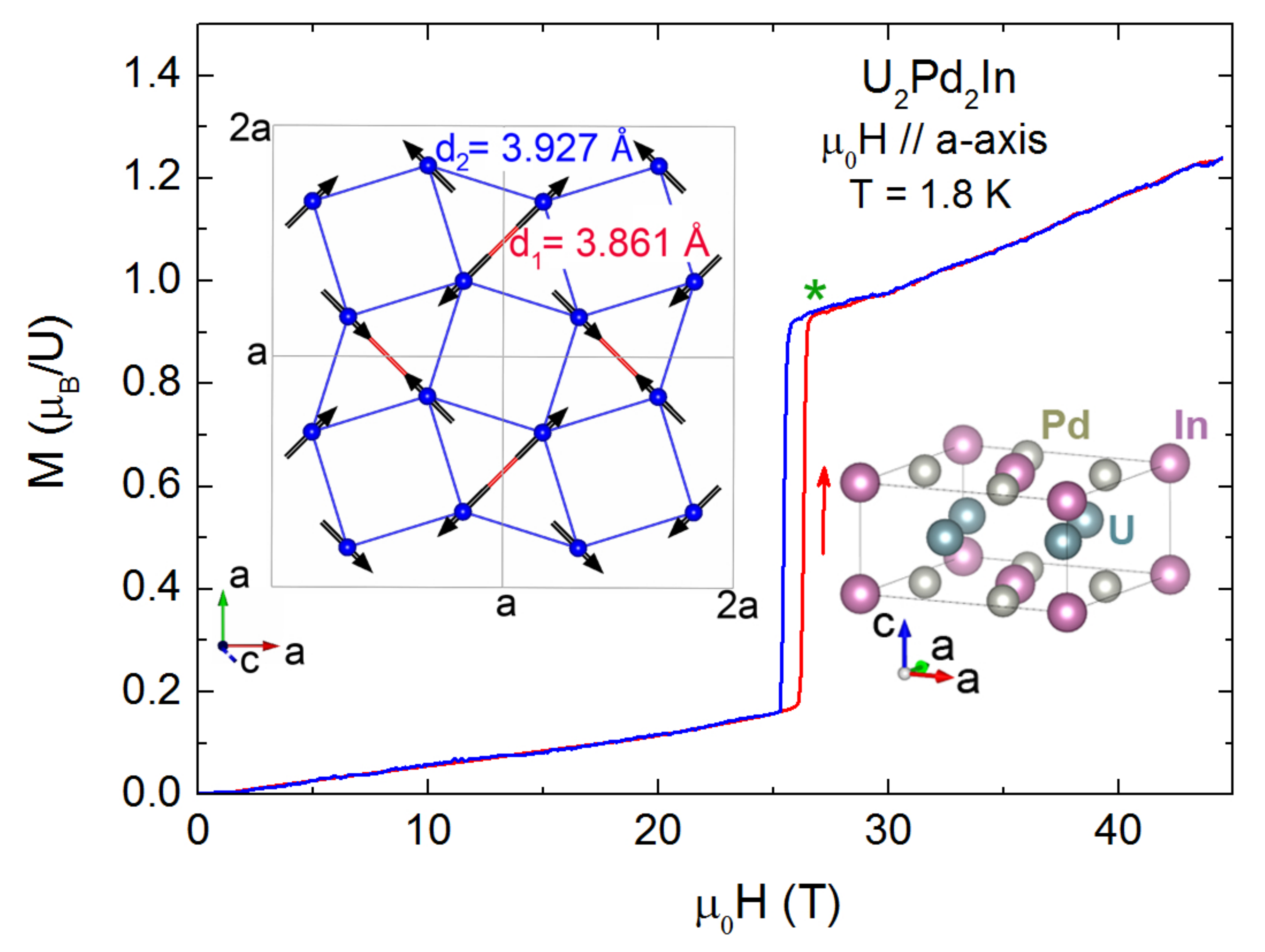}
\caption{(Color online) (a) Field dependence of the \UPI~single crystal magnetization measured at 1.8 K with pulsed field applied along the [100] crystallographic direction with increasing and decreasing magnetic field, respectively. Left inset shows the ground-state noncollinear antiferromagnetic structure of \UPI. Four magnetic unit cells projected along the $c$-axis with only U atoms are shown. Bonds connecting U atoms at two different distances, documenting the equivalency of the U sub-lattice with the Shastry-Sutherland lattice are given in red (shorter bond) and blue (longer bond) colors, respectively. The right inset shows the crystal structure of \UPI. The star denotes the field at which the neutron diffraction data used to determine the field-induced phase were collected.} \label{fig1}
\end{figure} 

It has been recognized only recently, that in tetragonal \utx~compounds (space group $P$4/$mbm$, Z = 2), U atoms build the SS lattice, equivalent to the SrCu$_{2}$(BO$_{3}$)$_{2}$ (see the left inset of Fig.~\ref{fig1}), too. However, the physics of these materials is by far more complex due to participation of 5$f$ electron states and consequently a presence of relativistic effects~\cite{UPI15,UPI6,UPI4}. \UPI~adopts a tetragonal crystal structure, consisting of two types of basal planes stacked alternately along the tetragonal $c$-axis as shown in the right inset of Fig.~\ref{fig1}.

 \UPI, as a member of the \utx~family of compounds,~orders antiferromagnetically at T$_{N}$ = 37~K~\cite{UPI1,UPI5,UPI10,UPI15,UPI9,UPI11,UPI12}, and previous neutron-diffraction studies revealed that U magnetic  moments of about 1.6 $\mu_{B}$/U form a noncollinear structure with U moments confined to the basal plane shown in the left inset of Fig.~\ref{fig1}~\cite{UPI9}. Relativistic $ab-initio$ calculations showed ~\cite{UPI4,UPI6,UPI53,UPI54} that such a structure is a consequence of relativistic effects and strong spin-orbit coupling (SOC) along with symmetry of the lattice. In addition, it has been shown that DMI plays an important role in selecting the ground-state, too~\cite{UPI53,UPI54}. The leading parameter is the single-ion anisotropy that locks U moments along the [110]-type directions. Moments can only be tilted away from this direction at a significant energy cost ~\cite{UPI53}. 
 
\begin{figure}
\includegraphics*[scale=0.26]{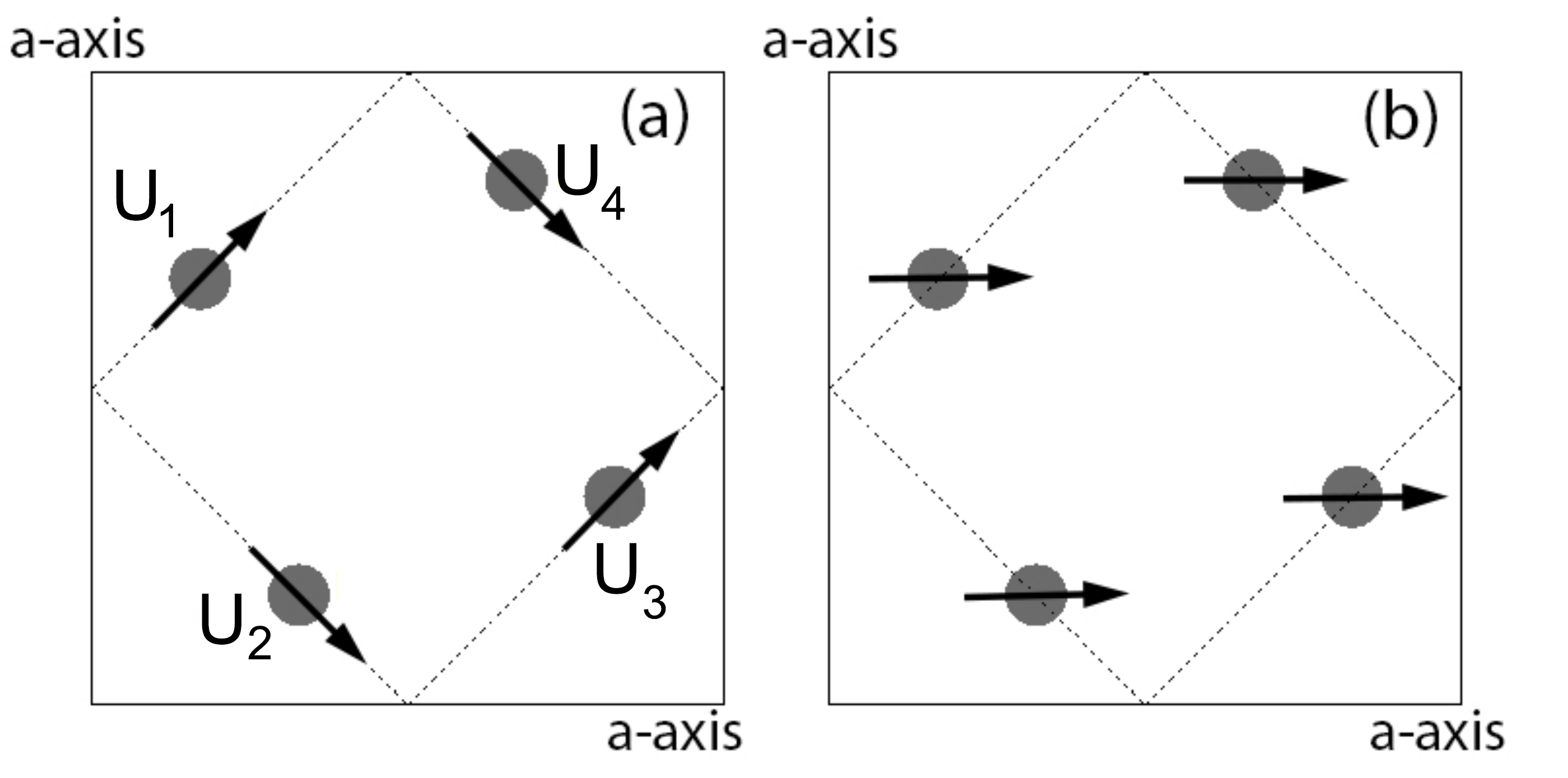}
\caption{(Color online) (a) Schematic representations of \UPI~magnetic structures predicted at low temperatures for fields above the critical value of $\approx$ 26 T applied along the [100] direction for a case with prevailing single-ion anisotropy (a) and for a case of a very strong  magnetic field (b), respectively.} 
\label{fig2}
\end{figure}

Here we report a surprising observation of a noncollinear field-induced magnetic structure in \UPI~that forms at low temperatures just above the critical field $\mu_{0} H_{c}$ = 25.8 T applied along the [100] direction. It appears that the structure is neither a noncollinear arrangement of U moments confined to the [110]-type directions expected in the case of prevailing single-ion anisotropy (see Fig.~\ref{fig2} (a)) nor a collinear ferromagnetic one shown in Fig.~\ref{fig2} (b) that is expected in the case of very strong magnetic field, respectively. Instead, a magnetic moment modulation described by two propagation vectors is observed. It can be qualitatively explained by a competition of single-ion anisotropy, DMI interaction that forces the moments to be orthogonal to each other and the application of magnetic field that forces the moment to align along its direction. To draw such conclusions, neutron diffraction data in fields up to 26 ~T are combined with static field bulk measurements up to 14 T and high-field magnetization measurements in pulsed fields up to 45 T. Our results are discussed in the context of recent first-principles calculations and provide insight into complex magnetization processes in the system.  

In our experimental studies, we used small crystals originating from the same batch as crystals used in previous studies~\cite{UPI11,UPI3,UPI7} with the mass of 11 mg (magnetization) and 48 mg (neutron diffraction), respectively. The magnetization $M$($T$) measurements in fields up to 45~T, generated by discharging a capacitor bank producing a 25~ms long pulse, were performed at the Hochfeld-Magnetlabor Dresden (HLD), Helmholtz-Zentrum Dresden-Rossendorf~\cite{UPI65}. Measurements were carried out at low temperatures with the field applied along the $a$-axis direction. The magnetic signal was detected using compensated pick-up coils and combined with magnetization and magnetic susceptibility measurements using the Quantum Design 14 T Physical Properties Measurements System (PPMS), which is part of the Laboratory for Magnetic Measurements at HZB. 

Neutron-diffraction data were collected at the HFM-EXED facillity~\cite{UPI2,UPI70,UPI16,UPI83} at the HZB. The crystal has been mounted on a copper cold finger of the $^{3}$He cryostat. We have investigated our \UPI~sample at the base temperature of 1.2 K, in fields up to 25.9 T applied along the $a$-axis direction. In order to increase the angular coverage and to reach desirable Bragg reflections, the magnet has been rotated, introducing an 11.85 degree angle between the field and neutron incident beam directions. The diffracted intensities used for refinements were typically collected for several hours with field kept constant.  

The data analysis was performed using Mantid~\cite{UPI19} software. Mantid software enables reliable corrections due to absorption $A$ and geometrical Lorentz factor $L$. For the absorption correction, the shape of the crystal was approximated by a sphere with a diameter of 1.2 mm. The geometrical Lorentz factor correction that depends  on the fourth power of the incident wavelength can be re-written to a form $L$ = $4d^{4}$ sin ($\theta$) that relates $L$ to the $d$ spacing of a reflection that enters also the Bragg law~\cite{UPI81}. The extinction correction has been neglected due to small size of the crystal and limited number of available reflections. For the evaluation of structure factors, the nuclear scattering lengths b(In) = 4.065 fm, b(Pd) = 5.910 fm and b(U) = 8.417 fm were used~\cite{UPI82} and for the analysis of magnetic intensities the standard U$^{3+}$ magnetic form factor was assumed.

Possible magnetic moment configurations allowed by the symmetry were derived using the representation analysis developed by Bertaut~\cite{Bertaut} and magnetic structure refinements were performed using Fullprof~\cite{Fullprof}.

In the main panel of Fig. \ref{fig1} we show magnetization curve obtained at 1.8 K for field applied along the [100] direction normalized to data obtained up to 14 T using PPMS. Both ascending and descending field branches exhibit a pronounced transition at $\mu_{0} H_{c}^{up}$ = 26.2 (1) T and $\mu_{0} H_{c}^{down}$ = 25.4 (1) T, respectively, leading to a significant hysteresis of about 0.8 T. The increase of the magnetization across the metamagnetic transition (MT) amounts to about 0.72 $\mu_{B}$/U. The MT is sharp and narrow at 1.8 K, getting smeared out as the temperature approaches T$_{N}$ = 37~K (not shown). We note that the magnetization above MT increases still significantly without saturation.

All these findings are in agreement with the literature, except the fact that the MT along the [100] direction is found to be lower and equal to critical field found for the [110] direction~\cite{UPI10,UPI15} pointing to incorrect identification of single crystal orientation in former experiments. In the present work we have microscopic evidence for the lower critical field being along the [100] direction using microscopic tool - the neutron diffraction in high static magnetic fields.

\begin{figure}
\includegraphics*[scale=0.4]{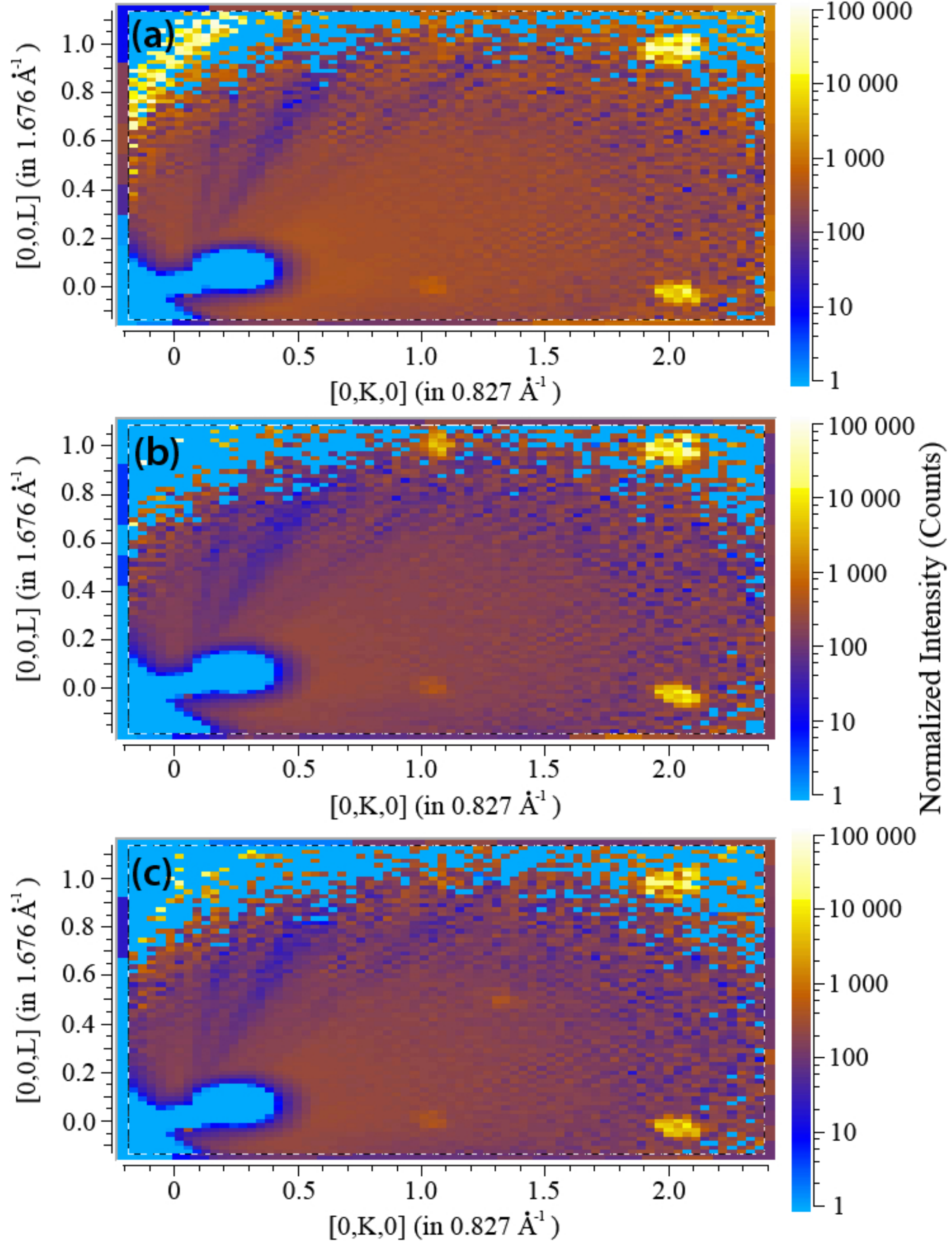}
\caption{(Color online) Portion of neutron diffraction data from the forward 
detector obtained on \UPI~single crystal at (a) 45 K in zero field, (b) at 1.2 K 
in zero field and (c) at 1.2 K in field of 25.9 T applied along the [100] 
direction converted to the reciprocal space with $h$ = 0.} \label{fig3}
\end{figure}

\begin{figure}
\includegraphics*[scale=0.45]{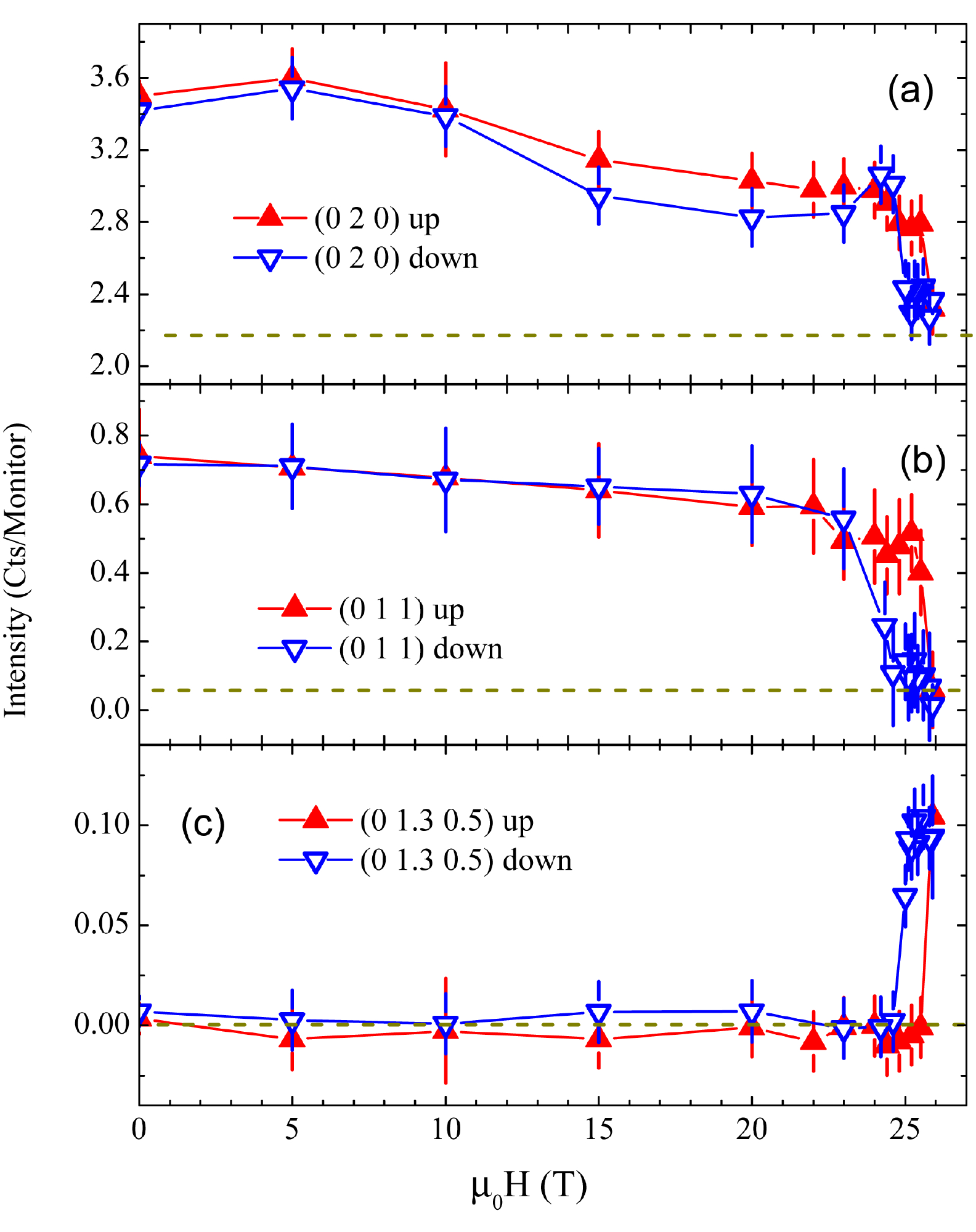}
\caption{(Color online) Field dependences of (a) (0 2 0), (b) (0 1 1) and (c) (0 1.30 $\frac{1}{2}$) Bragg reflections obtained by box-type integration of diffracted signal obtained with increasing and decreasing field applied along the $a$-axis direction at 1.2 K. Dashed line denote the intensity level at paramagnetic state.} \label{fig4}
\end{figure}

In Fig.~\ref{fig3} we show a portion of neutron diffraction data obtained at (a) 45 K in zero field and (b) at 1.2 K in zero field converted to the reciprocal space with $h$ = 0.  In total, the detector covers a reciprocal space range comprising six possible Bragg reflections with integer indexes: (0 1 0), (0 2 0), (0 3 0), (0 1 1), (0 2 1) and (0 3 1). In the paramagnetic state (panel (a)) we observe the (0 2 0) and (0 2 1) reflections that are in accordance with extinction rules for the space group $P$4/$mbm$. Integrated intensities of these two reflections are, after necessary corrections, in agreement with the intensities calculated from the crystal structure model of \UPI. However, at 45 K and zero field we observe also intensities at or near to the (0 1 0) and (0 1 1) reciprocal positions that are forbidden for the paramagnetic space group. One notes that these reflections are weak and temperature/field independent. In addition, they do not appear exactly at commensurate K=1 positions (see Fig.~\ref{fig3}) suggesting that they originate from a multiple scattering.  Indeed, azimuthal scans, i.e. scans around the scattering vector using a constant-wavelength instrument E4 at HZB showed that intensities of these reflections decrease with the azimuthal angle and vanish for rotation angle of about 8 degrees proving their multiple-scattering origin. 

In Fig.~\ref{fig3}(c) we show a portion of raw neutron diffraction data obtained at 1.2 K in field of 25.9 T applied along the [100] direction (i.e. \UPI~is above the MT, in the field-induced state) converted to the reciprocal space with $h$ = 0. Closer inspection and integration of intensities reveals that most of the Bragg reflections decrease in intensity upon increasing the field. Comparison with zero field data leads to a conclusion that Bragg reflections decrease to values very close to the intensities in the paramagnetic state. Merely the (0 2 1) and (0 2 0) reflections appear to contain still some magnetic intensity above MT. The magnetic intensities on top of these nuclear reflections are described by the propagation vector $q_{F}$ = (0 0 0). In Fig.~\ref{fig4} (a) and (b) we show field dependences of integrated intensities of (0 2 0) and (0 1 1) Bragg reflections normalized to the same monitor at 1.2 K, respectively. With increasing field both reflections decrease slowly up to the critical field where they drop abruptly to a level that is close to the intensity determined at 45 K zero field, denoted by the dashed line. With decreasing field a significant hysteresis is observed, which is in agreement with magnetization measurements. Irregularities seen around the MT, in particular higher intensity of the (0 2 0) Bragg reflection at the MT are ascribed to a field-induced extinction modifications.

\begin{figure}
\includegraphics*[scale=0.4]{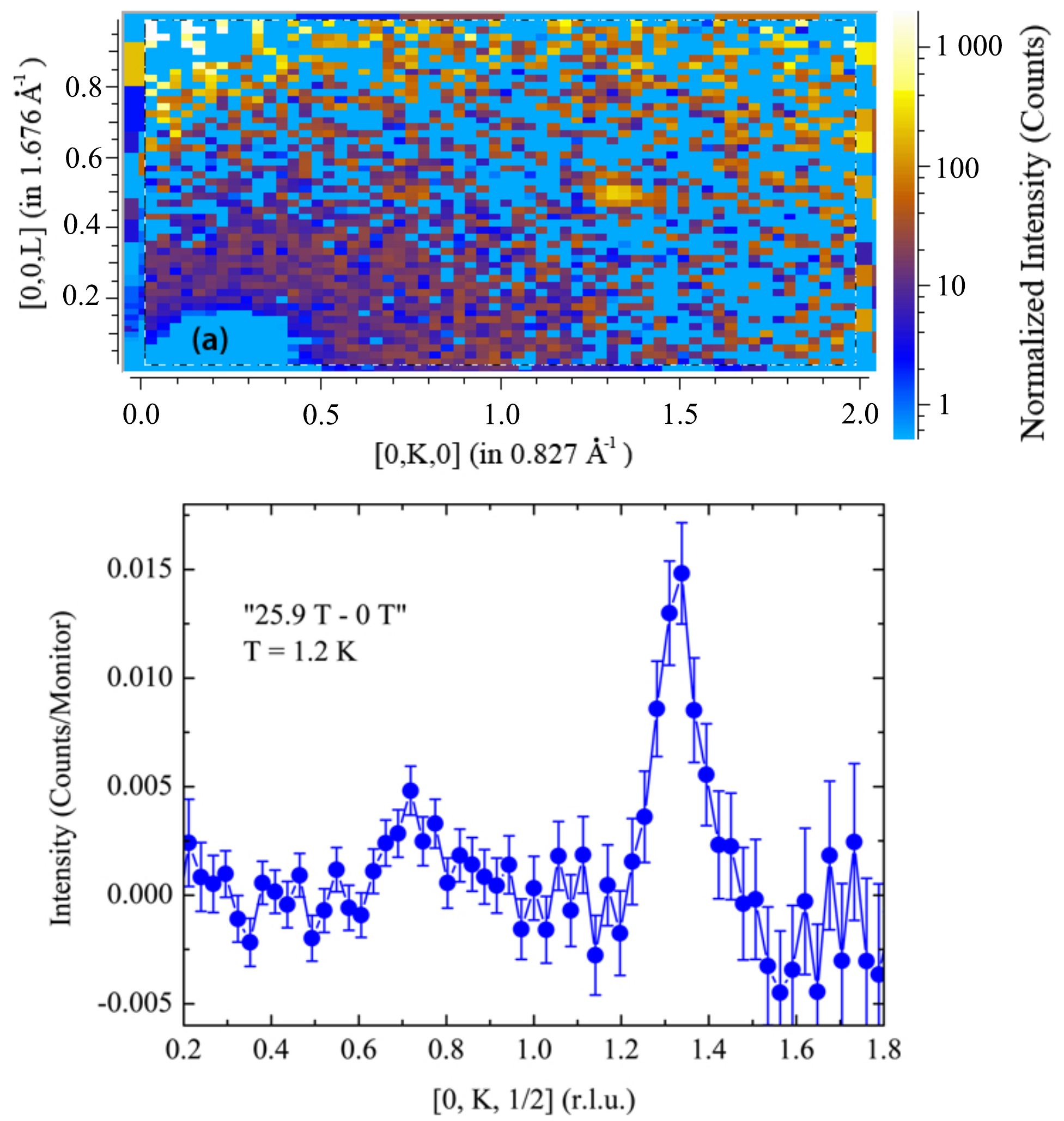}
\caption{(Color online) (a) Differential ``25.9 T - 0 T'' pattern of 
\UPI~obtained at 1.2 K and converted after normalization to the same monitor to 
the reciprocal space with $h$ = 0. Note two weak Bragg reflections indexable 
with (0 0.70 $\frac{1}{2}$) and (0 1.30 $\frac{1}{2}$) and absence of any 
intensity at (0 0.30 $\frac{1}{2}$) and (0 1.70 $\frac{1}{2}$). (b) projection of the data shown in (a) with 0.4 $\leq$ $L$ $\leq$ 0.6 on the K-index axis.} \label{fig5}
\end{figure}

A significant decrease of most of the Bragg reflections is in agreement with expected destruction of the ground-state antiferromagnetic structure and creation of a presumably ferromagnetic order in \UPI. However, we observe in this field-induced state surprisingly also Bragg reflections with fractional indexes. One of the reflections, indexable as (0 1.30 $\frac{1}{2}$), is clearly visible in panel (c) of Fig.~\ref{fig3}. Another one, indexable as (0 0.70 $\frac{1}{2}$), becomes apparent in the differential 25.9 T - 0 T pattern and its projection on the K-index axis (see Fig.~\ref{fig5}). We note that no (0 0.30 $\frac{1}{2}$) and no (0 1.70 $\frac{1}{2}$) reflections that are in the covered range, are visible. Also, we did not detect any higher-harmonic reflections suggesting that the modulation is of a pure sine-wave type. This observation suggests that the field-induced polarized state is not due to a simple ferromagnetic alignment of uranium moments but is more complex. The propagation vector of the modulation is $q_{AF}$ = (0 0.30(2) $\frac{1}{2}$). Here we denote this propagation vector with subscript AF to stress its antiferromagnetic character because it requires doubling along the $c$-axis direction. Let us note that in the small field interval above MT we do not observe any change in the propagation vector. 

Now we turn to the determination of the field-induced magnetic structure at 1.2 K  
in field of 25.9 T applied along the [100] direction, above MT. As mentioned, this state is characterized by two propagation vectors: the ferromagnetic one, $q_{F}$ = (0 0  0) and the antiferromagnetic vector $q_{AF}$ = (0 0.30 $\frac{1}{2}$). Possible 
magnetic moment configurations allowed by the symmetry were derived using the 
representation analysis developed by Bertaut~\cite{Bertaut}. This method, 
implemented in the computer code Basireps~\cite{Fullprof}, decomposes for a 
given vector the reducible paramagnetic group representation into a set of 
irreducible representations from which magnetic moment configurations at 
magnetic sites are deduced. The four U atoms within the tetragonal 
crystallographic unit cell of \UPI~occupy a single crystallographic site 4 $h$ 
($x$ $x$ $\frac{1}{2}$) with $x$ $\approx$ 0.172~\cite{UPI1,UPI5} denoted in 
Fig.~\ref{fig2} (a) as U(1) (at (0.172  0.672  0.500), U(2) at (0.328  0.172 0.500), 
U(3) at (0.828 0.328 0.500) and U(4) at (0.672 0.828 0.500). 
Analysis for the propagation vector  $q_{F}$ = (0 0 0) and the space group  
$P$4/$mbm$ has been performed earlier and is available in the 
literature~\cite{UPI51,UPI52}. It leads to models with U moments confined either 
within the basal plane or along the $c$-axis. Most of the possible 
configurations are antiferromagnetic and all four U moments appear to be 
connected by some symmetry operation. 

Only two of these models allow for ferromagnetic alignment of U moments, one 
with moments along the $c$-axis, the other one within the plane. Since the 
ground state magnetic structure is noncollinear with U moments within the basal 
plane and magnetic bulk measurements indicate an absence of MT for field applied 
along the tetragonal axis, we expect the moments to be oriented perpendicular to 
the $c$-axis. Indeed, the former model cannot account for observed magnetic 
intensities at (0 2 0) and (0 2 1) reflections at elevated fields. In 
particular, with the (0 2 0) reflection correctly reproduced, the latter 
reflection is calculated with too small intensity. 

The second possibility with moments confined within the basal plane stems from a 
two-dimensional irreducible representation that allows for a number of moment 
configurations. Here we reproduce the main results for this representation from 
Ref.~\cite{UPI52} (page 37), where it is denoted as $\Gamma _{10}$. In general, 
there are four independent coefficients (u, p, v, w) describing the magnetic 
structure: S(1) = (u+p, v+w, 0), S(2) = (u+p, v+w, 0), S(3) = (-u+p, v-w, 0) and 
S(4) = (-u+p, v-w, 0), where S(i) denotes the i-th magnetic moment listed above. 
These expressions can be rewritten to S(1) = ($\alpha$, $\beta$, 0), S(2) = 
($\alpha$, $\beta$, 0), S(3) = ($\gamma$, $\delta$, 0) and S(4) = ($\gamma$, 
$\delta$, 0)~\cite{UPI52}. Further, an assumption of moment equivalency leads to 
relations ($\alpha$ = $\pm$$\gamma$ and $\beta$ = $\pm$$\delta$) or ($\alpha$ = 
$\pm$$\delta$ and $\beta$ = $\pm$$\gamma$) and thus to two independent 
coefficients and a phase shift between them. In total, there are eight such 
possible combinations. 

\begin{figure}
\includegraphics*[scale=0.8]{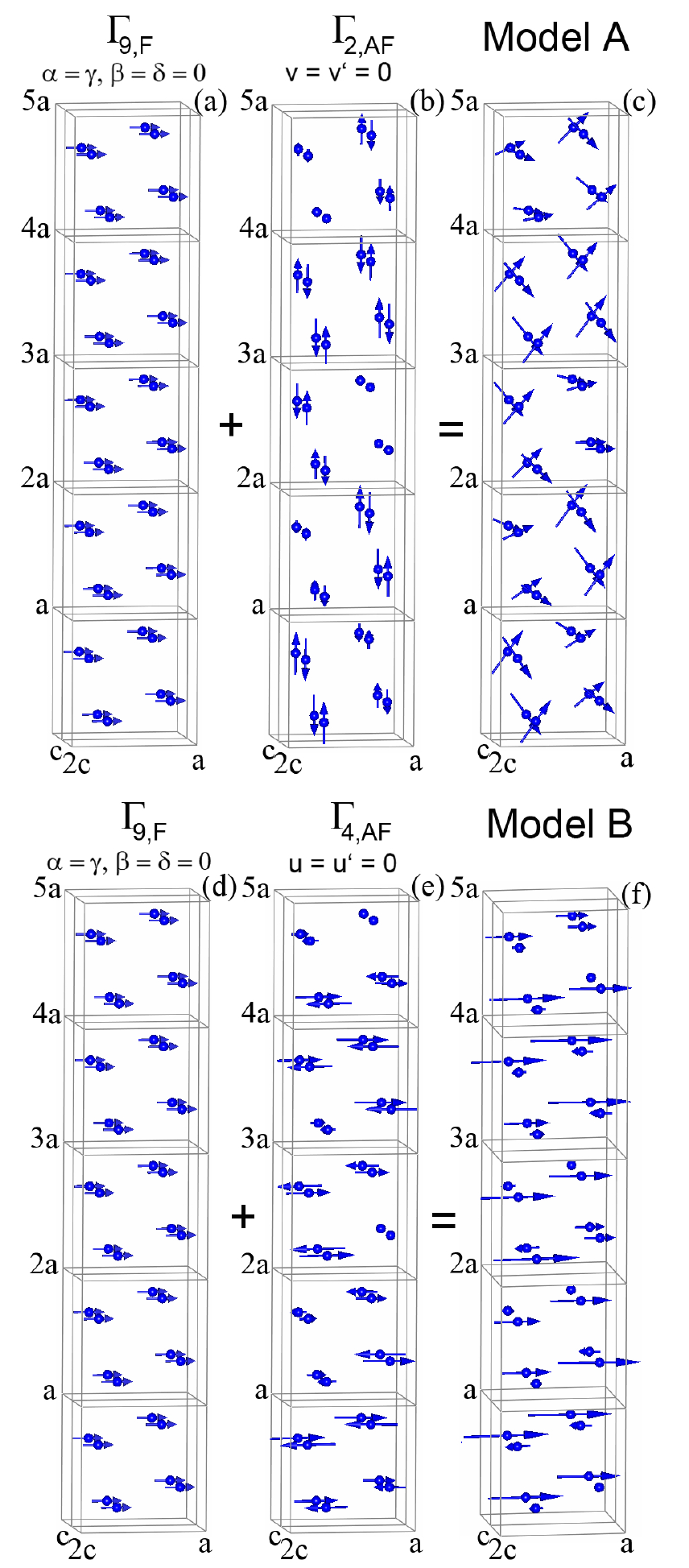}
\caption{(Color online) Possible magnetic structures belonging to the 
ferromagnetic $q_{F}$ = (0 0 0) (irrep. $\Gamma_{9}$)panel (a) and (d)) and antiferromagnetic 
$q_{AF}$ = (0 0.30 $\frac{1}{2}$) (irrep. $\Gamma_{2}$), panel (b) and irrep. $\Gamma_{4}$), panel (e)) propagation vectors, respectively. Resulting possible high-field induced magnetic structures combining the $q_{F}$ and $q_{AF}$ components are shown in panels 
(c) and (d) and denoted as Model A and Model B, respectively. Conditions for 
individual structures are written at the top of relevant structures (for details 
see the main text). In all panels $a$x5$a$x2$c$ crystallographic units are 
shown, containing only magnetic U moments.} \label{fig6}
\end{figure}

Among the allowed models, there are also configurations shown in Fig.~\ref{fig2} that were speculated about previously as possible 
field-induced magnetic structure in \UPI~at low temperatures~\cite{UPI3,UPI7}. 
The noncollinear model (a) leads to non-zero magnetic intensity at the (0 1 1) Bragg reflection. The other limiting model under consideration is collinear with all moments pointing along the $a$-axis as shown in Fig.~\ref{fig2} (b). It becomes apparent that the distinction between the two configurations depends on the existence/non-existence of a magnetic signal at the (0 1 1) Bragg reflection. Since we do not observe at 25.9 T any magnetic contribution on top of the (0 1 1) reflection, the magnetic configuration associated with ferromagnetic vector $q_{F}$ = (0 0 0) seems to be collinear, along the [100] direction shown in Fig.~\ref{fig2} (b). This collinear ferromagnetic arrangement is described by the irreducible representation $\Gamma _{9F}$ and this notation is used hereafter. Indeed, the refinement to all other allowed configurations lead to conclusion that the collinear arrangement shown in Fig.~\ref{fig6} (a) provides the best agreement between the calculated and observed intensities associated with $q_{F}$. The refined ferromagnetic component of uranium moment amounts to 0.6 (1) $\mu_{B}$. This value is only slightly lower than the magnetization step across the MT (see Fig.~\ref{fig1}), which amounts to 0.72 $\mu_{B}$/U.

The ferromagnetic component is, however, not the only component of the magnetic 
structure above MT. The other component comes from intensities associated with 
$q_{AF}$ = (0 0.30 $\frac{1}{2}$). The symmetry analysis made for $q_{AF}$ leads 
to a splitting of the four magnetic sites into two so-called orbits, each 
consisting of two magnetic sites. The first orbit comprises site U(1) and 
U(2), the other sites U(3) and U(4) as listed in Table~\ref{tab:irs}. There is no symmetry operation that would relate any of the sites in one orbit with any of the remaining site in the other orbit. Decomposition of the reducible paramagnetic representation for this propagation vector leads to four complex irreducible representations according to: $\Gamma _{m}$ = $\Gamma _{1}$  $\oplus$ 2$\Gamma _{2}$ $\oplus$ $\Gamma _{3}$ $\oplus$ 2$\Gamma _{4}$. Two representations, $\Gamma _{1}$ and $\Gamma _{3}$ lead to models with moments oriented along the $c$-axis, the remaining two allow for various couplings within the basal plane. They are listed in Table~\ref{tab:irs}. As can be seen, moments within one particular orbit are related but decoupled from moments in the other orbit. Moments between orbits may have further arbitrary phase shift.

\begin{table}[hb]
    \centering
    \caption{Possible magnetic moment symmetry relations for the four 
irreducible representations $\Gamma_{i}$ between U magnetic moments for the 
$P$4/$mbm$ space group and the magnetic propagation vector $q_{AF}$ = (0 0.30 
$\frac{1}{2}$) resulting from group theory. $\Gamma_{2}$ and $\Gamma_{4}$ allow 
for two basis vectors (BV). The four U atoms are situated at positions: U(1) (at 
(0.172  0.672  0.500) and U(2) at (0.328  0.172 0.500) (orbit one) and U(3) at 
(0.828 0.328  0.500) and U(4) at (0.672 0.828 0.500) (orbit two). $u$, $u'$, $v$ 
and $v'$ are magnetic Cartesian moment components and
$\alpha$* = exp$^{-2 \pi (1/2\delta)}$, 
where $\delta$ is the $y$ component of the propagation vector 
$q_{AF}$.}
    \label{tab:irs}
\begin{tabular}{cccccc}
\hline
\hline
$\Gamma_{i}$& BV & $\bold{S_k}$(U(1))   &   $\bold{S_k}$U(2) &      
$\bold{S_k}$U(3)   &   $\bold{S_k}$U(4)   \\

 \hline
$\Gamma_{1}$ &    BV1 &$0$ $0$ $u$ & $\alpha$*($0$ $0$ $u$) &  $0$ $0$ $u'$  & 
$\alpha$*($0$ $0$ $u'$)     \\
$\Gamma_{2}$ &    BV1 &$u$ $0$ $0$ & $\alpha$*($u$ $0$ $0$) &  $u'$ $0$ $0$ & 
$\alpha$*($u'$ $0$ $0$)  \\
 &   BV2 & $0$ $v$ $0$ & $\alpha$*($0$ $-v$ $0$) &  $0$ $v'$ $0$ & $\alpha$*($0$ 
$-v'$ $0$)  \\
$\Gamma_{3}$ &    BV1 &$0$ $0$ $u$ & $\alpha$*($0$ $0$ $-u$) &  $0$ $0$ $u'$ &  
$\alpha$*($0$ $0$ $-u'$)     \\
$\Gamma_{4}$ &    BV1 &$u$ $0$ $0$ & $\alpha$*($-u$ $0$ $0$) &  $u'$ $0$ $0$ & 
$\alpha$*($-u'$ $0$ $0$)  \\
 &   BV2 & $0$ $v$ $0$ & $\alpha$*($0$ $v$ $0$) &  $0$ $v'$ $0$ & $\alpha$*($0$ 
$v'$ $0$)  \\
\hline
\hline
\end{tabular}
\end{table}

Considerations analogous to those made for models associated with $q_{F}$ vector 
above lead to reduction of adjustable free parameters to three: Cartesian 
components along the $x$, along the $y$, being equal for the two orbits and a 
phase shift between the moments belonging to different orbits. A significant 
difference with respect to analysis given above is that the moments within a 
given orbit are not equal within a unit cell but related by a phase shift 
$\alpha$* = exp$^{-2 \pi (1/2\delta)}$ (see e.g. $\bold{S_k}$(U(1)) and 
$\bold{S_k}$(U(2)) in Table~\ref{tab:irs}) originating from the existence of the 
propagation vector  $q_{AF}$. After fitting the data to all eight possible 
moment configurations it appeared very clearly that four solutions give a 
good agreement with data. Let us note that we considered also the fact that 
reflections (0 0.3 $\frac{1}{2}$) and (0 1.7 $\frac{1}{2}$) have zero 
intensity at 25.9 T, which leads to a stability of fits and reduction of possibilities. 

While there is no good agreement found for models belonging to $\Gamma_{1}$ and $\Gamma_{3}$, there are two good solutions for $\Gamma_{2}$ and two for $\Gamma_{4}$. In the former case of $\Gamma_{2}$, all the U moments tend to have vanishing $x$ = [100] direction component, which is the direction along the applied field. Both these solutions form a collinear structure along the $y$ = 
[0 1 0] direction, i.e. they are modulated longitudinally. In Fig.~\ref{fig6}(b) we show one these solutions. Both possible solutions 
belonging to $\Gamma_{2}$ are of this type, they differ merely by the phase difference between U(1) and U(2) on one side and U(3) and U(4) on the other. However, they lead to substantially different U moment magnitudes of 1.65 (9) $\mu_{B}$ and 0.96 (6) $\mu_{B}$, respectively. The situation for the good models associated with the $\Gamma_{4}$ is similar except for the fact that for these solutions the U moments are oriented along the field direction, i.e. these magnetic structures have a vanishing $y$ component.
Again, the difference between them is merely the phase shift between the moments in the two orbits and the moment magnitudes that are 1.24 (7) $\mu_{B}$ and 0.66 (5) $\mu_{B}$, respectively. The former magnetic structure is shown in Fig.~\ref{fig6}(e). Common to all the magnetic structures connected with $q_{AF}$ is an antiferromagnetic coupling of U moments within one unit cell.

The resulting noncollinear magnetic structure of \UPI~at 1.2 K in magnetic field of 25.9 T applied along the [100] direction is determined by the two magnetic propagation vector components simultaneously. In Figs.~\ref{fig6}(c) and (f) we show two possible final high-field magnetic structures resulting from combinations of the best models obtained from the $q_{F}$ = (0 0 0) and $q_{AF}$ = (0 0.30 $\frac{1}{2}$) solutions, respectively. While the former model A consists of U moments making a complicated noncollinear arrangement with many moments nearly along (or perpendicular to) the [110]-type directions, the latter model B is collinear with moments along the field direction. We note that on the basis of the diffraction experiment alone it is impossible to conclude which of the magnetic structures is realized in \UPI~above the MT. However, it is reasonable to discard collinear solutions (see model B in Fig.~\ref{fig6} (f)) since such models lead to a very different U moments ranging from 0 to $\approx$ 2.0 $\mu_{B}$. Another reason is that the energy needed to turn all the moments out of the [110] type planes is enormous~\cite{UPI53,UPI54}.  

We conclude that the noncollinear solution shown in Fig.~\ref{fig6} (c), denoted as Model A is the correct one. In this model U moments vary between 0.7 and $\approx$ 2.4 $\mu_{B}$. The average moment amounts to  $\approx$ 1.6 $\mu_{B}$, i.e. value that is close to the ground-state value in contrast to the other solution connected with  $\Gamma_{2}$ that would lead to a much reduced value of $\approx$ 1.2 $\mu_{B}$/U.

As can be seen from Fig.~\ref{fig6}(c), the resulting high-field phase is incommensurate with U moments are predominantly within the [110] type planes or perpendicular to them. The $l$-component of the $q_{AF}$ causes the moments to be often perpendicular to each other as one moves along the $c$-axis direction from one unit cell to the adjacent one. In the direction perpendicular to the applied fields are U moments modulated incommensurately with the underlying crystal structure. A possibility of incommensurate magnetic structures in the SS system have been indicated by Chung $et$ al.~\cite{UPI58}. For certain ratio between the exchange along various bonds (see left inset of Fig.~\ref{fig1}) and spin magnitudes, incommensurate helical magnetic structures have been predicted to exist in zero field as the ground state. At this point we realize that our experimentally determined field-induced magnetic structure has an incommensurate sine-wave modulated component. Our restrictive geometry does not allow for the full identification/rejection of helices which are from the point of view of symmetry allowed as well but are not realized due to effect of applied field. However, we find it remarkable that we do find at elevated magnetic fields an incommensurate modulation of U moments. 

Let us now turn to the influence of the applied field. Field application introduces always additional unique axis into the problem. The number of symmetry operations reduces from original 16 operations of the paramagnetic space group $P$ 4/$m b m$ to 8 operations. For the field applied along the [100] direction all U moments remain mutually interconnected. This applies to a case, when the magnetic and crystallographic unit cells are of the same size. Experimentally determined propagation vector $q_{AF}$ = (0 0.30(2) $\frac{1}{2}$) suggests a larger magnetic unit cell, in particular modulation along direction perpendicular to the applied field and a doubling along the $c$-axis. The number of symmetry operations further decrease to 4: identity $E$, two-fold rotation axis $C_{2y}$ (+translation), the mirror plane $\sigma_{x}$  (+translation) and the mirror plane $\sigma_{z}$. The four moments split in two groups (orbits) and the doubling of the magnetic unit cell with respect to the crystallographic unit cell along the $c$-axis leads to antiparallel orientation of moments connected with the $\Gamma_{2,AF}$. The latter observation is in agreement with the instability of relativistic calculations assuming a the collinear ferromagnetic arrangement ~\cite{UPI53,UPI54}. 

The DMI vector is perpendicular to the nearest and the next-nearest U-U bonds, i.e. it is found along the $c$-axis. The applied field is perpendicular to the DMI vector and sudden changes in the propagation vector (and the magnetic structure) can be expected such as in the case of CsCuCl$_{3}$~\cite{UPI85} and Ba$_{2}$CuGe$_{2}$O$_{7}$~\cite{UPI84,UPI86}, where the zero-field helices get distorted for fields applied perpendicular to the unique axis leading to propagation vector modifications. In our case, one has to realize that there are no helices in \UPI~in zero field, where a noncollinear magnetic structure is found (see the left inset of Fig.~\ref{fig1}). Still, we do observe an appearance of a new propagation vector caused by reorientation of U magnetic moments.

The resulting magnetic structure can be qualitatively explained by competition of several different exchange interactions in the system. In zero field, the leading single-ion anisotropy keeps U moments aligned along the [110]-type directions. A perpendicular arrangement of U moments is supported further by the DMI that is in zero field 4-5 times weaker than the single ion anisotropy~\cite{UPI53,UPI54} and lifts also the degeneracy between two different orthogonal U moment arrangements~\cite{UPI53,UPI54}. The isotropic exchange interaction seems to be the weakest. The magnetic structure in magnetic fields is a consequence of a joint action of the three interactions mentioned above and the applied field that tries to align the U moments in a collinear fashion along the field direction. Further, our results document that one cannot always rely on the bulk magnetization results that indicate a pronounced magnetic phase transition but do not offer information regarding the microscopic arrangement of magnetic moments involved. Our results suggest that at least on the presently studied material (but applicable on many other anisotropic materials where a competition of different interactions occur) the field-induced state above MT is not a simple ferromagnetic phase. It is to be expected that a further increase of magnetic field turns U moments steadily towards the field direction, increasing the magnetization of the sample and explaining thus its significant increase above MT.


In conclusion, we report on neutron diffraction experiments on a Shastry-Sutherland model system \UPI~in DC magnetic fields up to 26~T combined with pulsed fields up to 45 T applied along the $a$-axis. The maximum available magnetic field in our neutron diffraction experiment is sufficient to cross the metamagnetic transition. The field-induced phase is noncollinear, described by two propagation vectors: a ferromagnetic one and an incommensurate AF one. The former vector corresponds to the collinear component of the structure that is aligned along the field direction, in agreement with the magnetization data. The AF vector describes a longitudinal modulation along direction that is perpendicular to the field direction. This vector also suggests an antiparallel coupling of moments in adjacent crystallographic units along the $c$-direction. The noncollinearity of the magnetic field-induced structure is a consequence of competition between the isotropic exchange, the single-ion anisotropy, DMI and the applied magnetic field. Our findings are in principle in agreement with the theoretical $ab-initio$ calculations~\cite{UPI53,UPI54}. However, the experimentally observed modulation of U moments for field applied along the [100] direction above the metamagnetic transition calls for further theoretical and experimental efforts. In particular, it appears, that the Hamiltonian used to explain the ground state magnetic structure does not capture all the physical details of the system and needs to be improved. 
 
\acknowledgments
We acknowledge the support of the HLD at HZDR, member of the European Magnetic Field Laboratory 
(EMFL).

\bibliography{UPI}

\end{document}